\documentclass[prb,superscriptaddress,twocolumn,floatfix]{revtex4}
\usepackage{graphicx}

\def\bscco{Bi$_2$Sr$_2$CaCu$_2$O$_{8+\delta}$}

\def\lbco{La$_{2-x}$Ba$_x$CuO$_4$}
\def\lbcoate{La$_{1.875}$Ba$_{0.125}$CuO$_4$}

\def\ybco{YBa$_2$Cu$_3$O$_{6+x}$}

\def\bscco{Bi$_2$Sr$_2$CaCu$_2$O$_{8+\delta}$}

\def\qaf{${\bf Q}_{\rm AF}$}

\begin{document}

\title{Testing the itinerancy of spin dynamics in superconducting \bscco}

\author{Guangyong Xu}
\author{G. D. Gu}
\author{M. H\"ucker}
\affiliation{Brookhaven National Laboratory, Upton, NY 11973-5000, USA} 
\author{B. Fauqu\'e}
\affiliation{Laboratoire L\'eon Brillouin, CEA-CNRS, CE-Saclay, 91191 Gif sur Yvette, France}
\author{T. G. Perring}
\affiliation{ISIS Facility, STFC Rutherford Appleton Laboratory, Chilton, Didcot, OX11 0QX, UK}
\affiliation{Department of Physics, University College London, Gower Street, London WC1E 6BT, UK}
\author{L. P. Regnault}
\affiliation{CEA-Grenoble, INAC-SPSMS-MDN, 17 rue des Martyrs, 38054 Grenoble Cedex 9, France}
\author{J. M. Tranquada}
\affiliation{Brookhaven National Laboratory, Upton, NY 11973-5000, USA} 
\date{\today}

\maketitle

{\bf
Much of what we know about the electronic states of high-temperature superconductors is due to photoemission\cite{lee07,yang08,kani08}  and scanning tunneling spectroscopy\cite{kohs08,gome07} studies of the compound \bscco. The demonstration of well-defined quasiparticles in the superconducting state has encouraged many theorists to apply the conventional theory of metals, Fermi-liquid theory, to the cuprates.\cite{lu92,liu95,bulu96,hwan07}  In particular, the spin excitations observed by neutron scattering at energies below twice the superconducting gap energy are commonly believed to correspond to an excitonic state involving itinerant electrons.\cite{esch06,fong99,he01,capo07,fauq07} Here, we present the first measurements of the magnetic spectral weight of optimally-doped \bscco\ in absolute units.  The lack of temperature dependence of the local spin susceptibility across the superconducting transition temperature, $T_c$, is incompatible with the itinerant calculations.  Alternatively, the magnetic excitations could be due to local moments, as the magnetic spectrum is similar to that in \lbcoate,\cite{tran04} where quasiparticles\cite{he09} and local moments\cite{huck08} coexist.
}

\bscco\ has been the cuprate system of choice for surface-sensitive techniques such as angle-resolved photoemission spectroscopy (ARPES) and scanning tunneling spectroscopy (STS) because it cleaves easily, thus allowing simple preparation of fresh surfaces.   ARPES\cite{lee07,yang08,kani08} and STS\cite{kohs08} studies of \bscco\ have convincingly demonstrated the existence of coherent electronic excitations (Bogoliubov quasiparticles) in the superconducting state. The excitation gap for quasiparticles has $d$-wave symmetry, going to zero at four nodal points along the nominal Fermi surface in the two-dimensional reciprocal space for a CuO$_2$ plane.\cite{lee07,yang08,kani08} On warming into the normal state, coherent electronic states are observed, at most, only over finite arcs about the nodal points; in the ``antinodal'' regions, there is a so-called pseudogap and an absence of quasiparticles.\cite{yang08,kani08}  

The ARPES results on \bscco\ have been used as a basis for predicting the magnetic excitation spectrum in the superconducting state, measurable by inelastic neutron scattering.\cite{esch06,fauq07}  To detect the magnetic signal with neutrons, however, one needs crystals of large volume in order to compensate for limited neutron source strength and weak scattering cross section, and such crystals have been difficult to grow.  The magnetic spectral weight is typically presented as the imaginary part of the dynamical spin susceptibility, $\chi''({\bf Q},\omega)$; here {\bf Q} is the wave vector of a magnetic excitation and $E=\hbar\omega$ is its energy, with $\hbar$ being Planck's constant divided by $2\pi$ and $\omega$ the angular frequency.  Previous neutron scattering studies\cite{fong99,he01,capo07,fauq07} of \bscco, working with crystals of limited size, focused on the change in $\chi''({\bf Q},\omega)$ on cooling through the superconducting transition temperature, $T_c$. 

A sustained effort has finally yielded crystals of sufficient size\cite{wen08} to allow a direct measurement of  $\chi''({\bf Q},\omega)$ on an absolute scale.  The as-grown crystals correspond to the condition of ``optimal'' doping (maximum $T_c$), with $T_c=91$~K.  To describe our results, we will follow previous practice\cite{fong99,he01,capo07,fauq07} and make use of a pseudo-tetragonal unit cell, with lattice parameters $a=3.82$~\AA\ (parallel to in-plane Cu-O bonds) and $c=30.8$~\AA.  Wave vectors {\bf Q} are given in reciprocal lattice units of $(2\pi/a,2\pi/a,2\pi/c)$.  The antiferromagnetic wave vector within a CuO$_2$ plane corresponds to ${\bf Q}_{\rm AF}=(\frac12,\frac12)$.  We note that the crystallographic unit cell is larger and orthorhombic, with a long-period incommensurate modulation in one in-plane direction.  The substantial atomic displacements associated with the incommensurability can modulate electronic and magnetic interactions\cite{slez08}; however, we did not detect any anisotropy of the magnetic scattering associated with the incommensurate modulation direction.

\begin{figure}[t]
\centerline{\includegraphics[width=3.6in]{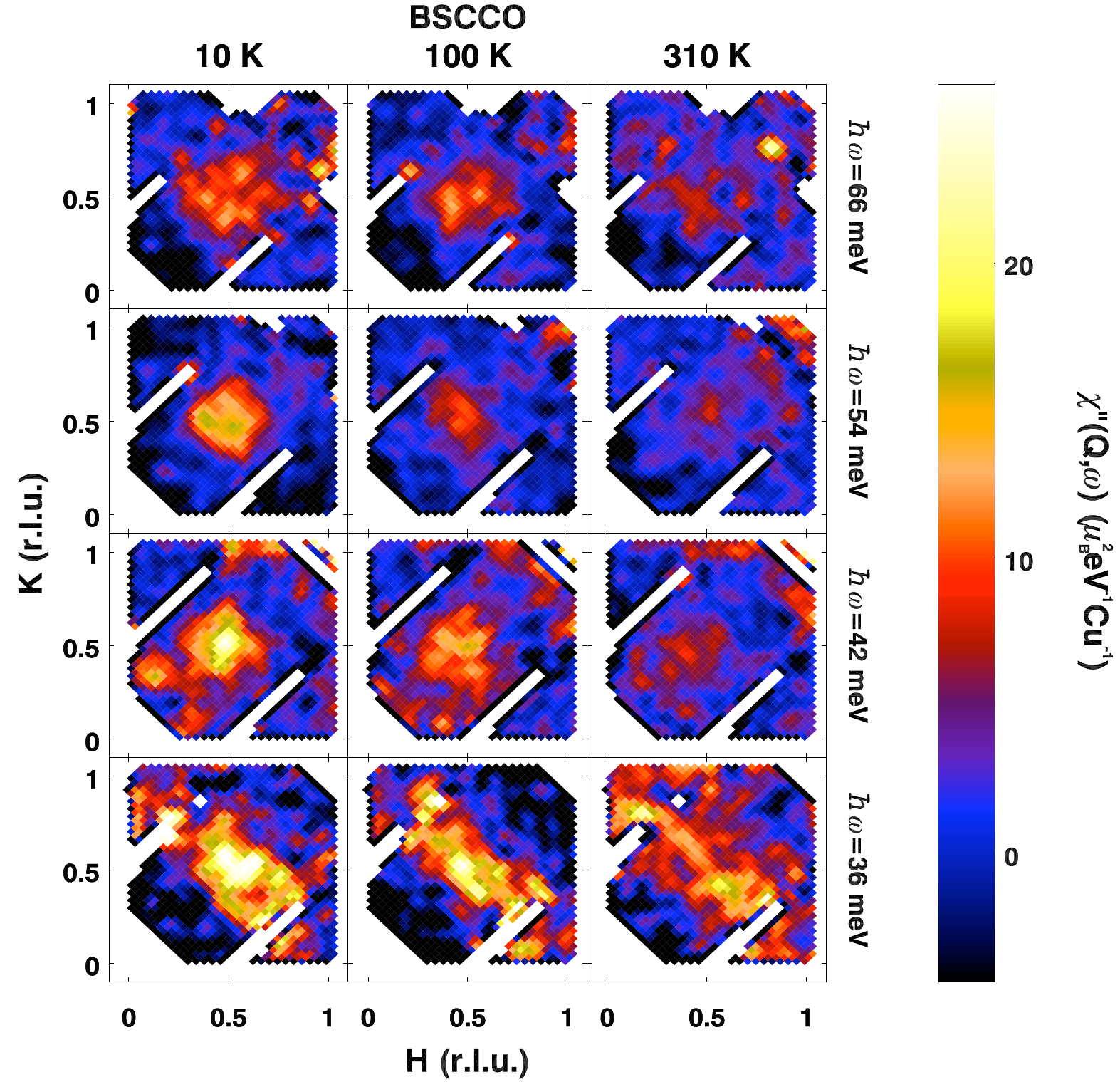}}
\caption{
{\bf  Constant energy slices through the scattered signal.} Time-of-flight data (for $E_i=120$~meV) converted to the form of $\chi''({\bf Q},\omega)$  (see Methods), with background subtracted (see text).  Data from four equivalent zones have been averaged to improve statistics.  Each column corresponds to a different temperature: 10~K, 100~K, and 310~K, from left to right.  Each row shows a different energy transfer: 36~meV, 42~meV, 54~meV, and 66~meV, from bottom to top.  For each slice, the data have been averaged over an energy range of $\pm3$ meV. White indicates areas not covered by detectors.
\label{fg:slices}}
\end{figure}

\begin{figure}[t]
\centerline{\includegraphics[width=3.2in]{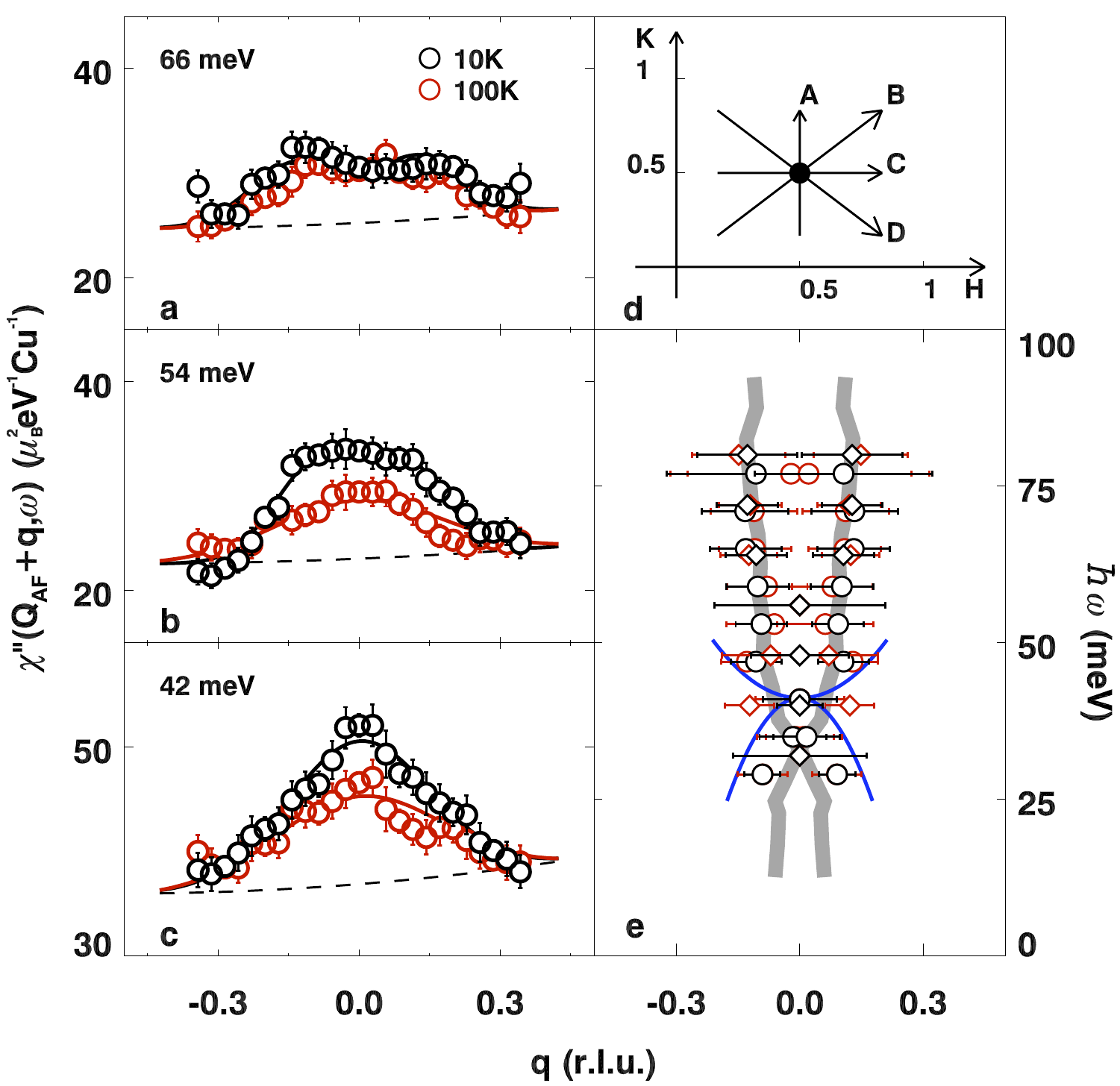}}
\caption{
{\bf Analysis of magnetic response.}
{\bf a-c}, Cuts through $\chi''({\bf q},\omega)$ (for $E_i=120$~meV) along ${\bf q}={\bf Q}-{\bf Q}_{\rm AF}$,  averaged over {\bf q} along  $[1,0,0]$ and $[0,1,0]$ (paths A and C, shown in {\bf d}).  Energy transfers are: {\bf a}, 66~meV; {\bf b}, 54~meV; {\bf c}, 42~meV.  Black symbols: $T=10$~K; red symbols: $T=100$~K;  error bars indicate standard deviations. The solid lines through the data represent fitted gaussian peaks---either a single peak or a pair of peaks constrained to be symmetric about ${\bf q}=0$.  Dashed lines indicate background.
{\bf e},  Plot of peak positions  obtained from the peak fitting.  Circles: $E_i=120$~meV; diamonds: $E_i=200$~meV; bars indicate peak widths.  For $\hbar\omega>40$~meV, points represent fits to the averaged A and C cuts; virtually identical results were obtained for B and D cuts.  For $\hbar\omega<40$~meV, only fits to D cuts (as in Fig.~\ref{fg:cuts2}) were used, because of the more complicated phonon background.  Solid blue lines indicate magnetic dispersion for YBa$_2$Cu$_3$O$_{6.95}$ from Reznik {\it et al.}\cite{rezn04};  gray lines indicate results for YBa$_2$Cu$_3$O$_{6.5}$ from Stock {\it et al}.\cite{stoc05}  
\label{fg:cuts} }
\end{figure}

The magnetic response was measured by inelastic neutron scattering using a time-of-flight (TOF) technique (see Methods).  Figure~\ref{fg:slices} presents constant-energy slices of the scattered signal as a function of momentum transfer about \qaf\ for several energies and temperatures.  The results are plotted in the form of $\chi''({\bf Q},\omega)$, as discussed in Methods.  Besides the magnetic scattering, peaked at \qaf, the measurements also include contributions from phonons; the phonon contribution should be approximately independent of temperature, but should vary, on average, as $Q^2$.  In this figure, we have subtracted a fitted background of the form $c_0+c_2Q^2$, where the coefficients $c_0$ and $c_2$ depend on energy, with $c_2$ independent of temperature.  (The magnitude of the background can be seen in Fig.~\ref{fg:cuts}a--c, where plots of the measured intensity along specific directions in {\bf Q} are plotted, with the fitted background indicated by a dashed line.)  The magnetic signal is expected to be weaker and more diffuse at 310~K, and the signal found there is generally consistent with this expectation.  The diagonal streak of scattering for $\hbar\omega=36$~meV at 310~K indicates a phonon contribution with intensity modulated in {\bf Q}.

\begin{figure}[t]
\centerline{\includegraphics[width=2.5in]{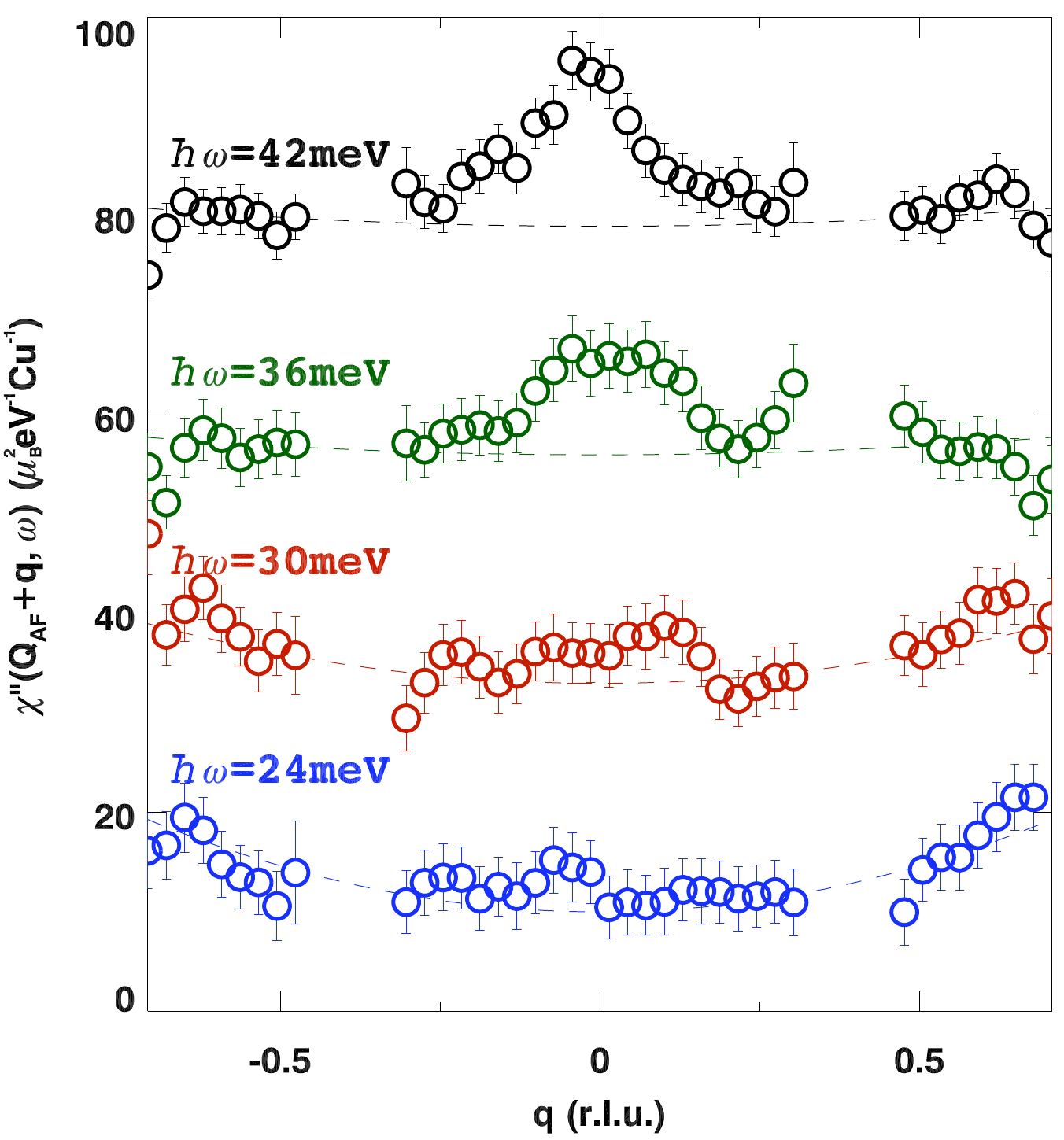}}
\caption{{\bf Magnetic response at lower energies}  Cuts through $\chi''({\bf q},\omega)$  along ${\bf q}=[1,-1,0]$ (path D in Fig.~\ref{fg:cuts}d) for excitation energies of 24, 30, 36, and 42 meV (bottom to top) measured at $T=10$~K with $E_i=120$~meV.  Data sets have been offset vertically for clarity.  Dashed lines indicate background.  Error bars indicate standard deviations.
\label{fg:cuts2}}
\end{figure}

For a more quantitative analysis of the magnetic signal and its temperature dependence, we can take cuts through these images and fit gaussian peaks to the structures.  Examples of such plots are shown in Fig.~\ref{fg:cuts}a-c and Fig.~\ref{fg:cuts2}.  From the fits, we obtain both the magnitude and {\bf Q} dependence of the magnetic response. The effective dispersion of the magnetic excitations is plotted in Fig.~\ref{fg:cuts}b.  Within the uncertainties, it is effectively isotropic about \qaf.  At low temperature, the overall shape of the dispersion is qualitatively consistent with the hour-glass spectrum that appears to be universal among the cuprates,\cite{birg06} with the excitations crossing \qaf\  at $\hbar\omega\sim40$~meV\cite{fauq07}; the response falls off below a spin gap energy of $\sim25$~meV (see Fig.~\ref{fg:cuts2}).  The effective upward dispersion, above 40~meV, is comparable to that observed\cite{stoc05} in underdoped YBa$_2$Cu$_3$O$_{6.5}$, and much steeper than that\cite{rezn04} in optimally-doped YBa$_2$Cu$_3$O$_{6.95}$.  

\begin{figure}[t]
\centerline{\includegraphics[angle=270,width=3.5in]{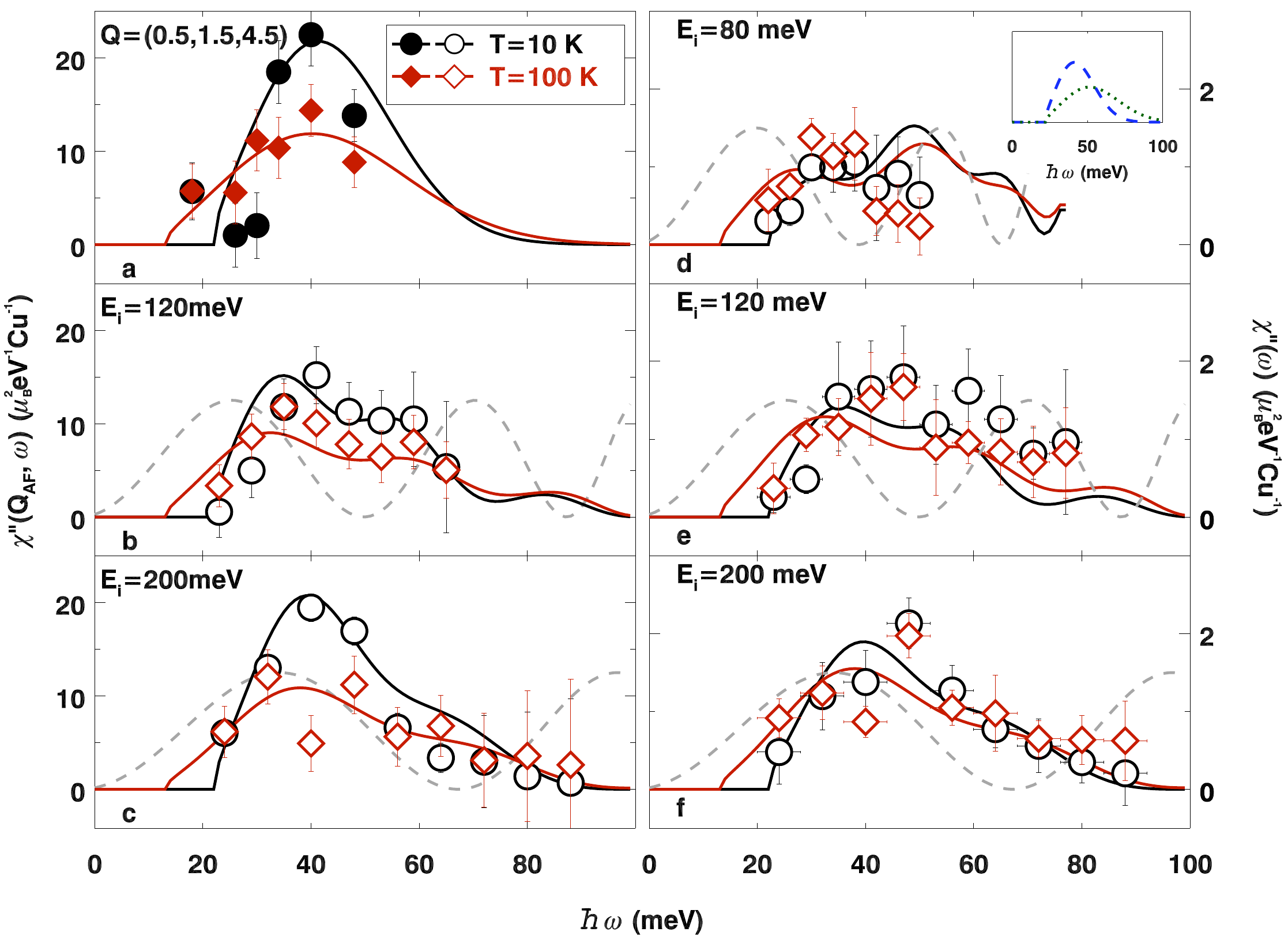}}
\caption{
{\bf Temperature and energy dependence of $\chi''$.}  
{\bf a--c}, $\chi''({\bf Q},\omega)$ at ${\bf Q}={\bf Q}_{\rm AF}$, with 
obtained from: {\bf a}, spin-polarized beam measurements at ${\bf Q}=(0.5,1.5,4.5)$; {\bf b}, TOF measurements with $E_i=120$~meV; {\bf c}, TOF data with $E_i=200$~meV.  Error bars indicate standard deviations.  {\bf d--f}, Local dynamic susceptibilty, $\chi''(\omega)$, for: {\bf d}, $E_i=80$~meV; {\bf e}, $E_i=120$~meV; {\bf f}, $E_i=200$~meV.  For all panels, black circles: $T=10$~K; red diamonds: $T=100$~K; vertical bars indicate standard deviations, based on the least-squares fits used to evaluate and integrate the {\bf Q} dependence of $\chi''(\omega,{\bf Q})$, while horizontal bars indicate energy range over which the signal was averaged.   Inset in {\bf d} shows a rough estimate of $\chi''(\omega)$ at 10~K for odd (dashed) and even (dotted) components; at 100~K, gap is decreased, widths increased, areas kept approximately constant.  These estimates are used (with amplitude adjustments for {\bf a--c}) to produce the solid lines in each panel, taking into account the weight variations  of the odd and even components with $\hbar\omega$ and $E_i$;  gray dashed lines show the relative weight of the odd component (magnitude oscillates between 0 and 1).  
\label{fg:qaf} 
\label{fg:int} }
\end{figure}

The method that we have used to identify the magnetic response and separate it  from the large phonon background is dependent on assumptions.  To get a direct measure of the magnetic response, we have performed a second experiment with spin polarized neutrons (see Methods and Supplementary Information).  Because of intensity limitations associated with this technique, our useful results are limited to measurements at \qaf; these are presented in Fig.~\ref{fg:qaf}a.   The lower panels compare results for  $\chi''({\bf Q}_{\rm AF},\omega)$ extracted from fits to the TOF data.  The data in Fig.~\ref{fg:qaf}a and c are generally consistent, with a substantial temperature-dependent change in signal at $\hbar\omega\sim40$~meV.  More modest changes with temperature are seen in Fig.~\ref{fg:qaf}b, which corresponds to measurements with a different incident neutron energy, $E_i$.

To understand the sensitivity to $E_i$, we have to consider the fact that CuO$_2$ planes in\linebreak \bscco\ come in correlated bilayers, just as in \ybco.  The magnetic response from a pair of layers can be separated into components with symmetry that is odd or even with respect to exchange of the layers.  The relative cross sections for the odd and even responses depend on the momentum transfer component along the $c$-axis, $Q_c$.   For the TOF measurements, $Q_c$ varies with $\hbar\omega$ in a fashion that is determined by the choice of $E_i$.\cite{stoc05}  The relative weight of the odd contribution as a function of $E_i$ and $\hbar\omega$ is indicated by the dashed gray lines in Fig.~\ref{fg:int}; the sum of the weights for the odd and even components at any given $\hbar\omega$ is equal to one.  For $E_i=200$~meV (Fig.~\ref{fg:int}c), the response at $\hbar\omega\sim40$~meV is mostly odd.  For the polarized-neutron study, all of the measurements were done at a fixed wave vector that maximizes the odd component.  For $E_i=120$~meV (Fig.~\ref{fg:int}b), there is a bigger contribution from the even component, which shows less change with temperature.\cite{capo07}

The enhancement of the odd component of $\chi''({\bf Q}_{\rm AF},\omega)$ in the superconducting state for $\hbar\omega\sim40$~meV is consistent with previous work\cite{fong99,fauq07}; however, our absolute measurements of the magnetic response also indicate significant weight in the normal state.  To get another perspective on the temperature dependence, we have integrated the fits to the magnetic response over the in-plane momentum transfers (assuming $\chi''({\bf Q},\omega)$ to be isotropic in ${\bf q}={\bf Q}-{\bf Q}_{\rm AF}$) to yield the local susceptibility, $\chi''(\omega)$, plotted in Fig.~\ref{fg:int}d--f.  As one can see, there are minimal changes in $\chi''(\omega)$ between 10~K and 100~K.  Thus, the changes in $\chi''({\bf Q}_{\rm AF},\omega)$ appear to be associated with a narrowing of the {\bf Q} dependence at low temperature.   

Our results cast considerable doubt on the common view of the $\sim40$-meV ``resonance'' mode as a spin-1 excitation of a pair of quasiparticles in the superconducting state,\cite{esch06} where the excitation is presumed to be resonant or excitonic because it occurs at an energy below twice the maximum superconducting $d$-wave gap energy ($2\Delta$).  From ARPES\cite{lee07} and STS\cite{kohs08} studies, we know that $2\Delta\approx80$~meV for optimally-doped \bscco.  It has been established that the superconducting gap decreases to zero at $T=T_c$ for wave vectors on the nodal arc.\cite{lee07}  A pseudogap remains at antinodal wave vectors for $T>T_c$, but this involves only a weak depression of the electronic spectral function at the Fermi energy.\cite{gome07}   Thus, if the magnetic excitations were due to quasiparticles, we should have seen dramatic changes in $\chi''(\omega)$ between 10~K and 100~K over the entire measured energy range\cite{rezn08}; clearly, our results are inconsistent with such a picture.  Similar inconsistencies for an electron-doped cuprate have been emphasized by Kr\"uger {\it et al.}\cite{krug07}  Furthermore, 
the strong temperature dependence of $\chi''(\omega)$ inferred from a conventional quasiparticle-based analysis of optical conductivity data\cite{hwan07} is inconsistent with our direct measurement, raising a challenge to that analysis.

The magnitude of $\chi''(\omega)$ in the energy range of 40--50 meV (Fig.~\ref{fg:int}d--f) is smaller than, but within a factor of two of, that for YBa$_2$Cu$_3$O$_{6.95}$,\cite{woo06} La$_{1.84}$Sr$_{0.16}$CuO$_4$,\cite{vign07} and La$_{1.875}$Ba$_{0.125}$CuO$_4$.\cite{tran04}  In the case of \lbco, it has been shown that the magnetism is dominated by local moments on Cu atoms.\cite{huck08}  Given the comparable magnitude of the magnetic response, universal dispersion, and the limited sensitivity to the presence of a superconducting gap, it seems likely that the spin fluctuations in \bscco\ are due primarily to local-moment magnetism.  A recent study of YBa$_2$Cu$_3$O$_{6.95}$ reached a similar conclusion.\cite{rezn08} 

La$_{1.875}$Ba$_{0.125}$CuO$_4$ may seem an unlikely case to compare with optimally-doped \bscco, as the former compound exhibits charge and spin stripe order.\cite{fuji04,huck08}  Nevertheless,  a recent ARPES study\cite{he09} has demonstrated the coexistence of nodal quasiparticles with the localized electronic moments of the spin stripes,\cite{huck08} and there is evidence for two-dimensional superconducting correlations, as well.\cite{li07}  Explaining such counterintuitive behavior remains a challenge for theorists.

\section*{Methods}

The \bscco\ crystals were grown at Brookhaven using the traveling-solvent floating-zone method.\cite{wen08}  Plate-like single crystals with sizes up to $50\times7.2\times7$~mm$^3$ were obtained.  Magnetic susceptibility measurements indicate an onset of diamagnetism at $T_c=91$~K.  Four crystals, with a total mass of 19 g, were mounted on aluminum supports and co-aligned for the experiment on the MAPS time-of-flight spectrometer at the ISIS spallation facility, Rutherford Appleton Laboratory.   The sample was mounted in a closed-cycle He refrigerator for temperature control, with the $c$-axis oriented along the incident beam direction.  The combinations of (incident neutron energy $E_i$, Fermi chopper frequency, typical integrated beam current) used for the measurements were (80 meV, 250 Hz, 5000 $\mu$A h), (120 meV, 350 Hz, 9000 $\mu$A h), and (200 meV, 500 Hz, 7000 $\mu$A h).  Measured differential cross sections, $d^2\sigma/d\Omega dE$, were converted to absolute units by normalization to measurements on standard vanadium foils.  The dynamic susceptibility was extracted using the formula
\begin{equation}
  {d^2\sigma\over d\Omega dE_f} = {k_f\over k_i} \,{1\over 1-e^{-\hbar\omega/k_B T}}\,
    f_{\rm Cu}^2({\bf Q}) \chi''({\bf Q},\omega),
\end{equation}
where $f_{\rm Cu}({\bf Q})$ is the anisotropic magnetic form factor for Cu.\cite{sham93}

A second set of 7 crystals, with a mass of 18.75 g, was aligned with x-rays at Brookhaven (collective mosaic $<2^\circ$) for the experiment on IN22 at the ILL.  The spectrometer is equipped with a vertically-focusing monochromator and horizontally-focusing analyzer, both made of Heusler alloy crystals.  The sample, oriented with $[100]$ and $[039]$ directions in the horizontal scattering plane, was mounted in an ILL He-flow cryostat, inside of Cryopadum, a system providing spherical polarimetry capabilities.\cite{regn04}  The neutron polarization flipping ratio measured on a strong Bragg peak was 16.  To obtain the inelastic magnetic scattering cross section, we used an appropriate combination of spin-flip intensities measured for three orthogonal neutron polarizations (see Supplementary Information) with a fixed final energy of 30.5 meV.   The results were normalized to the time-of-flight data through non-spin-flip measurements of phonons.



\bigskip

\noindent {\bf Acknowledgments}\ \  We gratefully acknowledge assistance from C. Stock, J. S. Wen, and Z. J. Xu, and a critical reading of the manuscript by S. A. Kivelson.  This work was supported by the Office of Science, U.S. Department of Energy under Contract No.\ DE-AC02-98CH10886.

\noindent {\bf Competing Interests}\ \ The authors declare that they have no
competing financial interests.

\noindent {\bf Author contributions}\ \ Sample preparation: G. D. G., M. H.; experiments: G. Y. X., B. F., T. G. P., L. P. R., J. M. T.; data analysis: G. Y. X., B. F.; paper writing: J. M. T., G. Y. X.

\noindent {\bf Correspondence}\ \  Correspondence and requests for materials
should be addressed to J.M.T.~(email: jtran@bnl.gov).


\vfill
\newpage

\section*{SUPPLEMENTARY INFORMATION}

\section*{Polarization analysis}

Let {\bf Q} (in the horizontal scattering plane) define the $x$
direction, and let $z$ be vertical; then $y$ is the orthogonal 
direction in the scattering plane.  Using Cryopad,\cite{regn04} it is possible to set
the initial or final polarization direction to be along any of these
directions.  (Note that the sample sits in a zero-field, magnetically-screened environment.)  For our problem, it is sufficient to consider only cases
where the initial and final polarizations are parallel.  

We can measure scattered intensities as a function of polarization
orientation ($xx$, $yy$, or $zz$).  If the flipper is off, then we will
detect neutrons that have the same spin orientation as in the incident
beam (designated $++$ or $--$); with the flipper on, we detect spins that
have been flipped by scattering from the sample ($+-$ or $-+$).  Thus, we
can label the measured intensities as $I_{\alpha\alpha}^{ss'}$,
where $\alpha = x,\,y,\,z$, and $s = +,\,-$.  

For the scattering cross section from the sample, we will use
$\sigma_{\alpha\alpha}^{ss'}$ as an abbreviation for the
differential cross section; note that it is labeled with the same
polarization orientation and spin-direction indices.  Let $N$ represent
the contribution due to nuclear scattering (including contributions from
the sample holder, cryostat, etc.), and let
$M_\alpha$ represent the magnetic contribution due to fluctuations along
the $\alpha$ direction.  (Note that $M_x$ is zero because those
fluctuations are parallel to {\bf Q}, and only the fluctuations
perpendicular to {\bf Q} have a finite cross section.)  For
\bscco, there should be no significant nuclear spin incoherent
scattering, so we will assume that the background $B$ depends on counting
time, but not on spin direction.  With these definitions, all of the
relevant cross sections can be written in a simple from:
\begin{eqnarray}
 \sigma_{xx}^{++}  & = &  \sigma_{xx}^{--}  =  N + B, \\
 \sigma_{xx}^{+-}  & = &  \sigma_{xx}^{-+}  =  M_y + M_z + B, \\
 \sigma_{yy}^{++} & = & \sigma_{yy}^{--} = N + M_y + B, \\
 \sigma_{yy}^{+-} & = & \sigma_{yy}^{-+} = M_z + B, \\
 \sigma_{zz}^{++} & = & \sigma_{zz}^{--} = N + M_z + B,\\
 \sigma_{zz}^{+-} & = & \sigma_{zz}^{-+} = M_y + B.
\end{eqnarray}
From the spin-flip (SF) cross sections alone, we can uniquely extract the
magnetic cross section with the combination
\begin{equation}
   M_y+M_z =  2\sigma_{xx}^{+-} - \sigma_{yy}^{+-} - \sigma_{zz}^{+-}.
\end{equation}
\null

Of course, the measured intensities $I_{\alpha\alpha}^{ss'}$ are
limited by the degree of polarization achieved and maintained by the
monochromator, analyzer, and the guide fields.  (We will assume that the
spin flipper performs perfectly.)  To collectively characterize the effective polarizing efficiency, one typically measures the flipping ratio $R$ by sitting on a Bragg peak and measuring
the ratio of the intensity with the flipper off to that with the flipper
on (say, $I_{xx}^{++}/I_{xx}^{+-}$).  By evaluating the measured intensities, taking account of the finite
beam polarizations, one finds that 
\begin{equation}
 M_y + M_z = {R+1\over R-1}\left(2I_{xx}^{+-} - I_{yy}^{+-} -
   I_{zz}^{+-} \right).  
\end{equation}
This is the formula that we used to extract the magnetic cross section from measured SF intensities for the three polarization directions.

\end{document}